\documentclass[DIV8,10pt,a4paper,headings=small]{scrartcl}
\setkomafont{pagehead}{\small} 
\usepackage{setspace}
\usepackage{abstract}

\usepackage[english]{babel}
\usepackage{times}
\usepackage[square,sort&compress,comma,numbers]{natbib}

\usepackage{amsmath,amssymb,amsfonts,mathtools,mathbbol,amsthm}

\usepackage[T1]{fontenc}

\usepackage[font=footnotesize]{caption}

\usepackage{enumitem}
\setlist{nolistsep,itemsep=\parskip,leftmargin=0pt,font=\it,style=unboxed,labelindent=\parindent}

\newtheorem{theorem}{Theorem}

\usepackage{color}
\usepackage[pdfpagelabels]{hyperref} 
\hypersetup{colorlinks=true}

\usepackage[rightcaption]{sidecap}
\usepackage{tikz}
\usetikzlibrary{calc,fit,fadings,positioning,backgrounds}
\newcommand{\convexpath}[2]{
[   
    create hullnodes/.code={
        \global\edef\namelist{#1}
        \foreach [count=\counter] \nodename in \namelist {
            \global\edef\numberofnodes{\counter}
            \node at (\nodename) [draw=none,name=hullnode\counter] {};
        }
        \node at (hullnode\numberofnodes) [name=hullnode0,draw=none] {};
        \pgfmathtruncatemacro\lastnumber{\numberofnodes+1}
        \node at (hullnode1) [name=hullnode\lastnumber,draw=none] {};
    },
    create hullnodes
]
($(hullnode1)!#2!-90:(hullnode0)$)
\foreach [
    evaluate=\currentnode as \previousnode using \currentnode-1,
    evaluate=\currentnode as \nextnode using \currentnode+1
    ] \currentnode in {1,...,\numberofnodes} {
-- ($(hullnode\currentnode)!#2!-90:(hullnode\previousnode)$)
  let \p1 = ($(hullnode\currentnode)!#2!-90:(hullnode\previousnode) - (hullnode\currentnode)$),
    \n1 = {atan2(\x1,\y1)},
    \p2 = ($(hullnode\currentnode)!#2!90:(hullnode\nextnode) - (hullnode\currentnode)$),
    \n2 = {atan2(\x2,\y2)},
    \n{delta} = {-Mod(\n1-\n2,360)}
  in 
    {arc [start angle=\n1, delta angle=\n{delta}, radius=#2]}
}
-- cycle
}

\newcommand{\id}{\mathrm{id}}
\DeclareMathOperator{\Tr}{\operatorname{Tr}}
\DeclareMathOperator{\supp}{\operatorname{supp}}
\renewcommand{\ker}{\operatorname{ker}}

\DeclareMathOperator{\cov}{\operatorname{cov}}
\newcommand{\dist}{\operatorname{d}}
\newcommand{\D}{\mathrm{d}}
\newcommand{\I}{\mathrm{i}}
\newcommand{\E}{\mathrm{e}}

\newcommand{\1}{\mathbb{1}}
\newcommand{\NN}{\mathbb{N}}
\newcommand{\RR}{\mathbb{R}}
\newcommand{\LL}{\mathbb{L}}
\newcommand{\C}{\mathbb{C}}
\newcommand{\mc}[1]{\mathcal{#1}}
\newcommand{\K}{\mc{K}}

\renewcommand{\H}{\mc{H}}
\DeclareMathOperator{\G}{\operatorname{\mc{G}}}
\newcommand{\nn}{\mc{Z}}

\newcommand{\B}{\mc{B}}
\newcommand{\A}{\mc{A}}
\newcommand{\mcS}{\mc{S}}

\newcommand{\ex}[2]{\left\langle #1 \right\rangle_{#2}}
\newcommand{\norm}[1]{\left\Vert #1\right\Vert}

\newcommand{\ket}[1]{\left.\left|{#1}\right.\right\rangle}

\newcommand{\braket}[2]{\left\langle #1 \middle| #2 \right\rangle}

\newcommand{\argdot}{ \,\cdot \; }
\newcommand{\ad}{^\dagger}

\newcommand{\kw}[1]{\frac{1}{#1}}

\newcommand{\sdim}{\mu}
\newcommand{\sdimconst}{M}
\newcommand{\Lad}{\mc{L}\ad}
\renewcommand{\L}{\mc{L}}
\newcommand{\Vset}{V}
\newcommand{\Eset}{E}
\newcommand{\compl}[1]{{\Vset\setminus #1}}

\newcommand{\T}[3][]{T_{\mc{#1\ad}}(#2,#3)}

\newcommand{\Tad}[3][]{\tau_{\mc{#1}}(#2,#3)}

\newcommand{\LRv}{\exp(1)b \nn}

\newcommand{\trunc}[2]{#1_{\upharpoonright \mathnormal{#2}}}

\makeatletter
\newcommand{\affiliation}[1]{\def\@affiliation {#1}}
\renewcommand{\maketitle}{%
  \thispagestyle{empty}

  \begin{center}
  \begin{spacing}{1.5}
    {\large\usekomafont{title} \noindent\@title}\vspace{2mm}
  \end{spacing}

  {\usekomafont{title} \noindent\@author}\vspace{2mm}\\
  {\small \noindent\@affiliation}\vspace{2mm}\\

  {\small \noindent\@date}
  \end{center}
  \vspace{10mm}
}

\newcommand\blfootnote[1]{%
  \begingroup
  \renewcommand\thefootnote{}\footnote{#1}%
  \addtocounter{footnote}{-1}%
  \endgroup
}

\addto\captionsenglish{%
}

\newcommand{\listkeys}{
  Heisenberg picture, 
  Lieb-Robinson bound, 
  spin lattice system, 
  spin lattice model, 
  dissipative systems, 
  ground state, 
  clustering of correlations, 
  decay of correlations, 
  quasi-locality
  classical simulation, 
  transport, 
  t-DMRG,
  local Liouvillian,
  local Hamiltonian, 
  thermodynamic limit, 
  quantum information, 
  review, 
  introduction, 
  Jordan-Wigner transform, 
  Lindblad equation, 
  area law
}

\hypersetup{
  pdftitle={Lieb-Robinson bounds and the simulation of time evolution of local observables in lattice systems},
  pdfauthor={M. Kliesch, C. Gogolin, J. Eisert},
  pdfsubject={Condensed Matter Physics, Quantum Physics},
  pdfkeywords={\listkeys}
  }
\makeatother

\begin{document}
\title{Lieb-Robinson bounds and the simulation of time evolution\\ of local observables in lattice systems}
\blfootnote{Appeared as a chapter in the Book~\cite{the_book}.}
\author{Martin Kliesch, Christian Gogolin, and Jens Eisert}
\affiliation{Dahlem Center for Complex Quantum Systems,\\ Freie Universit{\"a}t Berlin, 14195 Berlin, Germany}

\date{\today}
\maketitle

\begin{abstract}
This is an introductory text reviewing Lieb-Robinson bounds for open and closed quantum many-body systems. We introduce the Heisenberg picture for time-dependent local Liouvillians and state a Lieb-Robinson bound that gives rise to a maximum speed of propagation of correlations in many body systems of locally interacting spins and fermions. Finally, we discuss a number of important consequences concerning the simulation of time evolution and properties of ground states and stationary states.
\end{abstract}

\section{Introduction}
In lattice systems one might expect that, due to the locality of the interaction, there is some limit to the speed with which correlations can propagate.
Similar to the light cone in special relativity, there should be a space time cone, outside of which a local perturbation of such a system should not be able to influence any measurement in a significant way.
That this intuition can indeed be made rigorous was first shown by Elliott H.\ Lieb and Derek W.\ Robinson in a seminal work \cite{Lieb1972-28} in 1972.

Today, the term Lieb-Robinson bound generally refers to upper bounds on the speed of propagation of some measure of correlation.
Outside the space time cone defined by this speed, any signal is typically exponentially suppressed in the distance. 
The results of Lieb and Robinson, originally derived in the setting of translation invariant 1D spin systems with short range, or exponentially decaying interactions \cite{Lieb1972-28} have since been tightened \cite{Has04-LSM,NacSim09_review} and extended to more general graphs \cite{Has06_clustering_gs,Nachtergaele2009-286}
and to interactions decaying only polynomially with the distance, both, for spin systems \cite{Nachtergaele2006_124} and fermionic systems \cite{Has06_clustering_gs} (see also Ref.~\cite{Nachtergaele2010} for a review).
Lieb-Robinson bounds have been proven for Liouvillian dynamics first in Ref.~\cite{Poulin2010-104}, where Liouvillian dynamics is a generalization of Hamiltonian dynamics that can also capture the effect of a certain type of noise.
The bounds have recently been strengthened for a specific subclass of Liouvillians in Ref.~\cite{DesVer13} and have been generalized to time-dependent Liouvillian dynamics in Refs.~\cite{Nachtergaele2011-552,LRTrotter}.
Indeed, Lieb-Robinson bounds provide the basis for a wealth of statements in quantum many-body theory,
mostly as a mathematical proof tool, but also as an argument justifying numerical techniques. We will touch upon these implications and discuss the simulation of time evolution in more detail.

To keep the presentation both self-contained and concise, we mainly focus on Liouvillian dynamics as presented in Ref.~\cite{LRTrotter}.
The chapter is structured as follows: 
In the beginning, we introduce the setting and the necessary notation in Sect.~\ref{sec:settingsandnotation}. 
This includes in particular an introduction to Liouvillian dynamics in both the Schr\"{o}dinger and Heisenberg picture and a discussion of the relevant measures for approximation errors that are needed to state the Lieb-Robinson bound and their physical interpretation.
In the last part of Sect.~\ref{sec:settingsandnotation} we explain the setting of spin lattice systems.
Next, we state a general Lieb-Robinson bound in Sect.~\ref{sec:liebrobinsonbounds} and mention various consequences.
In particular, we explain the locality and simulability of time evolution in more detail in Sect.~\ref{sec:consequencesofliebrobinsonbounds}.
Finally, in Sect.~\ref{sec:jordanwignertransformation}, 
we state the Lieb-Robinson bound for fermions and introduce the Jordan-Wigner transform, which is a 
mapping between spin systems and fermionic systems.

\section{Setting and notation}
\label{sec:settingsandnotation}
In this section we introduce the necessary formalism to describe the dynamics of spin lattice systems evolving under local Liouvillian dynamics,
including local Hamiltonian dynamics as a special case.
While Hamiltonian time evolution describes the dynamics of closed systems, Liouvillian dynamics also captures the case of 
so-called open quantum systems \cite{Lin76}, which are systems coupled to memoryless ``baths''. 
Such couplings can be used to model Markovian ``noise'' 
perturbing the evolution of the system. 
The formalism and results discussed here partially address the problem of developing a better understanding of ``imperfect systems'' and, in particular, their time evolution (see also the chapter of Claude Le Bris).

\subsection{Schr\"{o}dinger and Heisenberg picture for time-dependent Liouvillians}
We start by introducing some notation and some basic mathematical facts. 
For some Hilbert space $\H$ of finite dimension $\dim(\H)$ let us denote the space of linear operators on $\H$ by $\B(\H)$. 
Together with the \emph{Hilbert-Schmidt} inner product, defined by 
$\langle A, B\rangle\coloneqq \Tr(A\ad B)$ for $A,B \in \B(\H)$, the space of operators $\B(\H)$ is also a Hilbert space.
Importantly, this defines the \emph{Hilbert-Schmidt adjoint} of a \emph{superoperator}. 
A superoperator is a linear map $T : \B(\H) \to \B(\H)$, i.e., $T \in \B(\B(\H))$ and its (Hilbert-Schmidt) adjoint $T\ad \in \B(\B(\H))$ is defined via 
$\langle X, T\ad (Y)\rangle \coloneqq \langle T(X), Y \rangle$ for all $X,Y\in \B(\H)$.
The subspace of \emph{observables} $\A(\H) \subset \B(\H)$ are the Hermitian, i.e. self-adjoint operators and the set of \emph{states} $\mcS(\H)$ (also called density operators) are positive semidefinite Hermitian operators with unit trace. 
Given an observable $A \in \A(\H)$ and a state $\rho \in \mcS(\H)$ the expectation value is 
\begin{equation}
 \ex A \rho \coloneqq \Tr(\rho A). 
\end{equation}

When considering time evolution one is confronted with the following scenario: At some time $s$ the system is in some initial state $\rho$ and at a later time 
\begin{equation}
  t\geq s \qquad \text{(throughout this chapter)} 
\end{equation}
one measures some observable $A$ that gives rise to an expectation value $\ex A {\rho} (s,t)$. 
The time evolution can be described either in the Schr\"{o}dinger picture or the Heisenberg picture.
In the Schr\"{o}dinger picture, one evolves the initial state $\rho$, given at time $s$, 
forward in time until time $t$ is reached at which the measurement is performed. 
In the Heisenberg picture, in turn, one evolves the observable $A$ backwards in time from $t$ to the time $s$ at which the initial state is given.

In the Schr\"{o}dinger picture one considers the states to be time-dependent.
In the case of a closed quantum system evolving under a Hamiltonian $H$, 
the state of the system at time $t$ is the solution of the linear initial value problem
\begin{equation}
  \frac{\D}{\D t} \rho_s(t) =-\I [H(t), \rho_s(t)] , \quad \rho_s(s) = \rho ,
\end{equation} 
where the solutions of the dynamical equations carry the initial time $s$ as a label for reasons that become clear once we switch to the Heisenberg picture.
If a system is coupled to further degrees of freedom giving rise to decoherence and dissipation, 
one can, e.g., for many physically relevant situations with weak coupling, 
describe the system as an \emph{open quantum system} whose dynamic is given by the solution of the linear initial value problem 
\begin{equation}\label{eq:ME}
  \frac{\D}{\D t} \rho_s(t) = \Lad_t(\rho_s(t)) \ , \quad \rho_s(s) = \rho ,
\end{equation} 
where $\Lad : \RR \to \B(\B(\H))$ is called the \emph{Liouvillian}\footnote{As we will later mostly work in the Heisenberg picture it is convenient to denote the Liouvillian in the Schr\"{o}dinger picture by $\Lad$ rather than $\L$.}, and where the time dependence is given by the input $t\in\RR$. 
The Liouvillian may explicitly depend on time, e.g. to be able to capture change of external control parameters. Throughout this chapter we restrict the time dependence to be \emph{piecewise continuous}.
For an equation of motion of this form, the only constraint is that the time evolution maps states to states, i.e., is completely positive and trace preserving. This is equivalent \cite{WolCir08} to the Liouvillian $\Lad_t$ having a Lindblad representation \cite{Lin76}, i.e. it must be of the form
\begin{align}\label{lindbladform}
\Lad(\rho) = -\I [H, \rho]+ \sum_{\mu=1}^{\dim(\H)^{2}} \left(
    2 L_{\mu} \rho L_\mu \ad - L_\mu \ad L_\mu \rho - \rho L_\mu \ad L_\mu  \ \right),
\end{align}
for some time-dependent operators 
$H: \RR \to \mc A(\H)$ and $L_\mu: \RR \to \B(\H)$. 

Liouvillian dynamics is ubiquitous in many contexts in physics. 
It has recently been studied particularly intensely in the context of cold atoms in optical lattices 
\cite{Diehl,fermionic_atoms_opt_latt,Unravelling,Lesanovsky}, 
trapped ions \cite{Barreiro2011-470,Schindler_trapped_ions}, 
driven dissipative Rydberg gases \cite{Rydberg}, 
and macroscopic atomic ensembles \cite{Polzik}.
Also dissipative state preparation \cite{Cirac}, dissipative phase transitions \cite{Diehl}, noise-driven criticality \cite{Prosen1} and nonequilibrium topological phase transitions \cite{Topology}
have been considered.

The initial value problem \eqref{eq:ME} defines the \emph{propagator} (also called dynamical map) $\T[L]{t}{s}: \B(\H) \to \B(\H)$ via
\begin{equation}
 \T[L]{t}{s} (\rho) \coloneqq \rho_s(t),
\end{equation}
which is also the unique solution of the initial value problem
\begin{align}
 \frac{\D}{\D t} \T[L]{t}{s} = \Lad_t \T[L]{t}{s} \ , \quad \T[L] s s = \id. 
\end{align}
The expectation value at time $t$ then is
\begin{align}
\ex A {\rho} (s,t) = \Tr\left[\T[L]{t}{s} (\rho)\, A\right] .
\end{align}
If the Liouvillian $\Lad$ is time-independent, a state satisfying $\Lad(\rho)=0$ is called stationary state.  
The role played by stationary states is reminiscent of the role of 
ground states of Hamiltonians. For the case of a unique stationary state the spectral gap of the Liouvillian is a measure of the speed of convergence \cite{Kastoryano2013} towards this stationary state. 

Evolving some state $\rho$ from $s$ to $r \geq s$ and then from $r$ to $t\geq r$ also yields $\rho_s(t)$ and hence the propagator has the \emph{composition property} 
$\T[L] t r \T[L] r s = \T[L]{t}{s}$ for all $t\geq r\geq s$. For classical processes this property is stated by the \emph{Chapman-Kolmogorov equation}.
It is a good exercise to derive the differential equation
\begin{equation} \label{eq:Tbackward}
 \frac{\D}{\D s} \T[L]{t}{s} = - \T[L]{t}{s} \Lad_s ,
\end{equation}
from this property.

We are now ready to introduce the Heisenberg picture, in which the states are constant and the observables are defined as solutions of a dynamical equation.
Of course, both pictures must yield the same expectation values, i.e., 
\begin{equation}
 \ex A \rho (s,t) = \Tr\left(\rho\, \Tad[L] t s (A) \right), 
\end{equation}
where 
\begin{equation}
\Tad[L]st = \T[L]{t}{s}\ad
\end{equation}
is the adjoint of $\T[L]ts$ in the Hilbert-Schmidt inner product. 
$\tau_{\mc L}$ is the \emph{propagator in the Heisenberg picture}. Using Eq.~\eqref{eq:Tbackward}, it is not hard to see that it is the unique solution of
\begin{equation}
 \frac{\D}{\D s} \Tad[L]st = - \L_s \Tad[L]st  \ , \quad \Tad[L]tt = \id ,
\end{equation}
where $\L$ and $\Lad$ are Hilbert-Schmidt adjoints of each other and, in particular, $\L$ is given by
\begin{equation}
 \L (A) = \I [ H, A] + \sum_{\mu=1}^{\dim(\H)^{2}} \left(
    2 L_{\mu}\ad A L_\mu  - L_\mu \ad L_\mu A - A L_\mu \ad L_\mu  \ \right) .
\end{equation}
Now we define the \emph{(backward) time evolved} observable $A_t(s)$ to be 
the solution of 
\begin{equation}
 \frac{\D}{\D s} A_t(s) = \L_s(A_t(s)) \ , \quad A_t(t) = A ,
\end{equation}
which is equivalent to  
\begin{equation}
 A_t(s) = \Tad[L] ts(A).
\end{equation}
In the case of time-independent Liouvillians, one can equivalently define the Heisenberg picture such that observables are evolved forward in time. 
More generally, this is always possible if
$\Tad[L] st  \L_t = \L_t \Tad[L] st $ for all $s \leq t$,
i.e., when the propagator commutes with the Liouvillian. 
In this case, one can equivalently evolve observables forward in time with $\T[L] ts \ad$ which is then $\T[L] ts \ad=T_{\L}(t,s)$. 
If the propagator and the Liouvillian do not commute, there is no simple way to obtain a consistent forward time evolution for $A$. 

\subsection{The physically relevant norms}
Norms are functions that quantify the ``size'' of a vector or operator and hence provide an important tool to measure errors when approximating observables. Let us explain this in more detail.
The Hilbert space inner product induces a norm via 
$ \norm{\ket \psi} \coloneqq \sqrt{\braket \psi \psi} $. 
This norm gives rise to a norm on operators:  
let $B \in \B(\H)$, then its \emph{operator norm} is defined to be the supremum
\begin{equation}
 \norm{B} \coloneqq \sup_{\norm{\ket \psi} =1} \norm{B \ket \psi},
\end{equation}
which coincides with the largest singular values of $B$.
If $B$ is an observable, then its norm is its largest eigenvalue in magnitude and thus a bound on the range of values one can obtain when $B$ is measured, i.e.,
\begin{equation}
 \norm{B} = \sup_{\rho \in \mcS(\H)} |\Tr(\rho B)| .
\end{equation}
Considering the case where $B=A-A'$ is the difference of two observables $A, A' \in \A(\H)$ this means that the operator norm is the physically relevant norm to measure closeness of the two observables:
If $\norm{A - A'}$ is small, then $A$ and $A'$ will have almost the same expectation value on all states, see Ref.~\cite{NieChu00} for a more detailed discussion.

\subsection{Lattice systems and local Liouvillians}
\begin{figure}[tb]
  \centering
\begin{tikzpicture}[scale = .8, 
    site/.style = {circle, draw = black, very thick, fill = gray, inner sep = .5ex},
    edgeline/.style = {draw = black},
    edgefill/.style = {blue!20}
    ]
     \newcommand{\drawedge}[2]{
        \draw [edgeline] \convexpath{#1}{#2};
	\begin{scope}[on background layer]
	 \fill [edgefill] \convexpath{#1}{#2};
	\end{scope}
    }
    \begin{scope}[transform shape]
    \foreach \x in {0,...,4}
      \foreach \y in {0,...,3}
	\node [site] (n\x\y) at (\x,\y){};
    \draw [edgeline] (n40.center) circle (.3cm);
    \end{scope}
    \begin{scope}[on background layer]
	\fill [edgefill] (n40.center) circle (.3cm);
    \end{scope}
    \foreach \X in {{n11,n22},
		    {n12,n21},
		    {n32,n33,n43},
		    {n30,n31,n41},
		    {n01,n11,n10},
		    {n02,n03,n13},
		    }{
	\drawedge{\X}{.3cm}
     }
     \foreach \X in {{n23,n33},
		      {n10,n20},
		      {n31,n22},
		      {n32,n42}}{
	\drawedge\X {.25cm}
     }
     \foreach \X in {{n20,n30},
		      {n42,n41},
		      {n23,n22}}{
	\drawedge \X {.275cm}
     }
     \path (n31.center) ++ (.45,-.45) node (X1){$X$}
	   (n41.center) ++ (0,.5)     node (X2){$Y_1$}
	   (n31.center) ++ (-.5,.5)   node (X3){$Y_2$}
	   (n30.center) ++ (-.5,0)    node (X4){$Y_3$};
\end{tikzpicture}
  \caption{An interaction hypergraph. 
	   The dots denote the vertices and the frames the hyperedges. 
	   The maximum number of nearest neighbors is $\nn = 4$: 
	   the edge $X$ has the nearest neighbors $Y_j$ and itself.}
  \label{fig:lattice}
\end{figure}
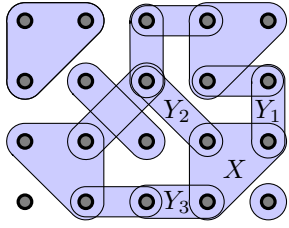
Quantum lattice systems are formally described by a set of (spatial) sites that are considered to be the vertices of a (hyper)graph. The interactions between the sites correspond to the edges of the (hyper)graph (see also Fig.~\ref{fig:lattice}).
In this section we explain this setting for spin systems in detail and consider fermionic systems in Sect.~\ref{sec:jordanwignertransformation}. 

Let us assume that the set of sites $\Vset$ is finite and that each site $x \in \Vset$ is associated with a finite dimensional Hilbert space $\H_x$.
The Hilbert space of some subsystem $X \subset \Vset$ is denoted by
$\H_X \coloneqq \bigotimes_{x \in X}\H_x$ and 
$\H \coloneqq \H_{\Vset}$.
For an operator $A \in \B(\H)$ we define its support $\supp(A)$ to be the smallest subset $X \subset \Vset$ such that it acts as the identity outside of $X$, i.e., 
$A_X = A \otimes \1_{\compl{X}}$. 
The set of operators supported on $X$ is denoted by 
$\B_X(\H) \coloneqq \{A \in \B(\H): \supp(A) \subset X\}$ and the subspace of observables by $\A_X(\H) \subset \B_X(\H)$.
For a Liouvillian $\L$ on $\B(\H)$ we define its support to be 
\begin{equation}
 \supp(\L) \coloneqq \bigcup \{X \subset \Vset : \A_{\compl{X}}(\H) \subset \ker(\L) \},
\end{equation}
i.e., the part of the system where $\L$ corresponds to a non-trivial time evolution. 
The set of Liouvillians supported on $X$ is denoted by $\LL_X(\H)$. 
Often we omit the Hilbert space and write, e.g., $\A_X$ instead of $\A_X(\H)$.

We are interested in the time evolution under \emph{local Liouvillians}. A Liouvillian $\L$ is called \emph{local} if it is of the form
\begin{equation}
 \L = \sum_{X \subset \Vset} \L_X,  \quad  \L_X \in \LL_X .
\end{equation}
In many physically relevant situations many of the strictly local terms $\L_X$, in particular those belonging to large sets $X$, will be zero.
This structure reflects interactions and dissipation processes that are finite-ranged.
The \emph{interaction graph} $\Eset$ of the Liouvillian is the set of all subsets of $\Vset$ for which the Liouvillian contains a non zero term, i.e.,
\begin{equation}\label{eq:Esetdef}
  \Eset \coloneqq \{ X \subset \Vset : \L_X \neq 0 \} .
\end{equation}
As an example, consider the case of a 1D system with nearest neighbor interactions and open boundary conditions.
If the sites are $\Vset=\{1,\hdots,N\}$, the interaction graph is $\Eset = \{\{1,2\},\{2,3\},\hdots,\{N-1,N\}\}$ in that case.

The interaction (hyper)graph $\Eset$ defines a distance $\dist(X,Y)$ between any two sets $X,Y \subset \Vset$ of vertices.
The distance $\dist(X,Y)$ is equal to $0$ if and only if $X \cap Y \neq \emptyset$ and otherwise equal to the length of the shortest path connecting $X$ and $Y$, and $\infty$ if there is no connecting path.
A path between two sets $X,Y \subset \Vset$ is a sequence of elements of $\Eset$, such that the first element contains a vertex in $X$, each element of the path shares at least one vertex with the following element and the last element contains a vertex in $Y$. Note that $d$ is a degenerate metric on subsets of $\Vset$. 
In the above 1D example the graph distance of the two sets $\{j\},\{k\} \subset V$ would simply be 
$\dist(\{j\},\{k\}) = |j-k|$, as one would expect.

\section{A Lieb-Robinson bound}
\label{sec:liebrobinsonbounds}
In this section we state and explain a very general Lieb-Robinson bound for the speed of propagation of correlations in spin systems under arbitrary time-dependent Liouvillian dynamics.
Our goal is to make statements about \emph{local} time evolution, i.e., time evolution of local observables arising from local interactions and local noise.
In order to make this precise, let us impose some technical constraints on a possibly time-dependent Liouvillian $\L_s \in \LL_{\Vset}$, which we consider to be fixed from now on. 
Local time evolution is captured by a Liouvillian $\L$ that is a sum of strictly local terms $\L_{X}$, each of which is bounded in norm by $b$, and a maximum number of nearest neighbors $\nn$. 
In more detail, we define
\begin{align}\label{eq:localLdef}
 \L &= \sum_{X \in \Eset} \L_X,  \quad  \L_X : \RR \to \LL_X(\H) , \text{ piecewise continuous},
 \\
 \label{eq:bdef}
 b &\coloneqq \sup_{s, X}\norm{\L_X(s)}, 
 \\
 \label{eq:nndef}
 \nn &\coloneqq \max_{X\in \Eset}
    \left|\left\{Y \in \Eset : Y \cap X \neq \emptyset \right\} \right|.
\end{align}
The parameters $b$ and $\nn$ will determine the Lieb-Robinson speed and also the final results about the \emph{spatial truncation}
\begin{equation}
\trunc{\L}{\Vset'} \coloneqq \sum_{X \subset \Vset'} \L_X 
\end{equation}
of the Liouvillian $\L$ to some region $\Vset' \subset \Vset$. 
Now we are ready to state the Lieb-Robinson bound for this setting.
Similar results on Liouvillians can be found in Refs.~\cite{Nachtergaele2011-552,Poulin2010-104}.
The theorem is quite general and it might not be immediately obvious how statements about propagation of information are implied. But this will become clear in the next section.
\begin{theorem}[Lieb-Robinson Bound \cite{LRTrotter}\footnote{
  In Ref.~\cite{LRTrotter} the bound is given for an arbitrary metric on the vertex set and the Liouvillians are allowed to have interaction range $a$ in that metric. 
  Our interaction graph distance $d$ is induced by a metric on $\Vset$ for which $a=1$.}] \label{thm:LR}
Let $\L: \RR \to \LL(\H)$ be a local Liouvillian as specified in Eqn.~\eqref{eq:localLdef} --  \eqref{eq:nndef} and $X,Y \subset \Vset$. 
Then, for every $\K_Y \in \LL_Y(\H)$, $A_X \in \B_X(\H)$, and $s\leq t$
\begin{equation}\label{eq:LRbound}
 \norm{\K_Y \Tad[L]s t(A_X)} 
 \leq C \norm{\K_Y} \norm{A_X} \E^{v (t-s) - \dist(X,Y)},
\end{equation}
where $v = \LRv$ and $C$ is some constant depending polynomially on the size of the smaller of the two sets $X$ and $Y$. 
\end{theorem}
Remembering that the Liouvillian maps an observable to its time derivative. The theorem tells us that an evolved observable $\Tad[L] s t (A_X)$ remains basically unchanged when evolved with respect to a Liouvillian $\K_Y$ that is supported on a region a distance much larger than $v (t - s)$ away from $X$, i.e., that $\Tad[L] s t (A_X)$ is almost the identity outside the corresponding space-time cone.
More intuitively, the Lieb-Robinson bound tells us that information travels with a velocity bounded by the \emph{Lieb-Robinson speed} $v$ of the considered lattice system.
In the special case $\K_Y = i[B_Y , \argdot]$ for some $B_Y \in \A_Y(\H)$, 
Eq.~\eqref{eq:LRbound} yields a Lieb-Robinson bound in the more common form of an upper bound on the commutator $\norm{[B_Y, \Tad[L] s t(A_X)]}$ (compare Refs.~\cite{Nachtergaele2011-552,Poulin2010-104}).

If a system is mixing in the sense that all states are driven towards a steady state then information encoded in the initial state gets lost at some point. This puts an upper bound on the distance over which information can propagate. Therefore, one might expect that there is some effective Lieb-Robinson speed that decreases in time. This is indeed true for certain systems with fluctuating disorder \cite{BurEisOsb08} and for a certain class of Liouvillian dynamics \cite{DesVer13}.

Finally let us mention that, the lattice can also be infinitely large (implied by the next theorem), but the restriction to finite-dimensional subsystems is not merely for simplicity of notation: 
For infinite-dimensional systems the situation can be quite different. For some anharmonic lattices \cite{Nachtergaele2009-286},
and other instances of strongly correlated models \cite{Schuch} Lieb-Robinson bounds can still be found, 
as well as for commutator-bounded operators \cite{commutator_bounded_LR}. 
Still, counterexamples to Lieb-Robinson bounds are known for models with infinite-dimensional constituents \cite{supersonic}.

\section{Consequences of Lieb-Robinson bounds}
\label{sec:consequencesofliebrobinsonbounds}
Lieb-Robinson bounds are fundamental for a plethora of statements concerning various properties of locally interacting systems. 
We first discuss immediate consequences as far as the dynamics of such systems is concerned. 
Next, we turn to implications for the classical simulation of time evolution. Finally, we discuss static properties that can be derived from Lieb-Robinson bounds.

\subsection{Quasi-locality of quantum dynamics}
\label{sec:quasilocality}
The result of the last section suggests that the terms of the Liouvillian whose support is sufficiently far away from the support of an observable are irrelevant for the time evolution. 
More precisely, one should be able to spatially truncate the Liouvillian $\L$ to some region $\Vset' \subset V$.
If $X$ is sufficiently far from the boundary of $\Vset'$, i.e., if $\dist(X, \compl{\Vset'})$ is larger than the radius $v\cdot (t-s)$ of the space time cone of $\Tad[L] s t(A_X)$, then the dynamics of $A_X$ under the truncated Liouvillian $\trunc{\L}{V'}$ and the original Liouvillian $\L$ should be very similar.
In the next theorem we will see that this is indeed the case if the underlying interaction graph is of finite spacial dimension, which we define first. Let us denote the ``sphere'' around some subsystem $X\in \Eset$ with radius $n$ by 
\begin{equation}
 S_X(n) \coloneqq \{ Y \in \Eset : \dist(Y,X) = n\}.
\end{equation}
Then we say that an interaction graph $\Eset$ is of \emph{spatial dimension} $\sdim$ if there is a constant $\sdimconst >0$ that only depends on local properties of the interaction graph 
such that for all $X \in \Eset$
\begin{equation}\label{eq:sdimdef}
\left| S_X(n) \right| \leq M n^{\sdim-1} .
\end{equation}

For example, the interaction graph of next-neighbor Liouvillians on a $\mu$-dimen\-sional cubic lattice has dimension $\mu$.

\begin{figure}[tb]
  \centering
  \colorlet{conecol}{gray!40}
  \colorlet{conecolwhite}{gray!1}
  \colorlet{Vcol}{green!60!black}
  \colorlet{Acol}{blue}
  \begin{tikzpicture}[scale=.75,
    node distance = .8ex,
    axis/.style = {->},
    tick/.style = {color=gray},
    site/.style = {circle, draw, fill=black, inner sep = .3ex, anchor=center},
    obs/.style = {transform shape,rectangle, draw=black, fill = Acol!50, opacity=0.8},
    Vprime/.style = {transform shape, draw, minimum height = .5ex, inner ysep=0, inner xsep=.5ex, fill=Vcol!20},
    helplines/.style = {very thin, dashed},
    widtharrow/.style = {very thin, <->}
    ]
    \def \slen{.05}
    \def \nysteps{7}
    \def \ticksdist{.5}
    \def \nsites{17}
    \def \sitedist{.5}
    \def \shadewidth{.8}
    \def \angle{30}
    \def \tanangle{0.57735}
    \def \xclip{3}
    \def \yclip{3.5}
    \def \shadexwidth{0.92376} 
    \def \Dmarkdist{2ex}
    \newcommand{\pointcoordinate}[1]{\draw [<-, thin, red] (#1) -- ++(.5,.5) node [anchor = south west] {#1};}

    \foreach \x in {0,...,\nsites}
    \node (s\x) at ($(\x*\sitedist,0)$){};
    \node (V) [Vprime, fit = (s2) (s15)]{};
    \draw [axis] (-0.5,0)  -- (9,0);
    \foreach \x in {0,...,\nsites}
    \filldraw (s\x.center) circle (.4ex);
    \node (innerX)[obs, fit = (s8) (s9)]{};
    \node (X)[transform shape,fit = (innerX), inner xsep=0, inner ysep=.5ex] {};
    \path (X.north east) ++($(\tanangle*\yclip, \yclip )$) coordinate (cre);
    \path (X.north west) ++($(-\tanangle*\yclip, \yclip )$) coordinate (cli);
    \coordinate (origin) at ($(X.north)-(X.center)$);
    \begin{scope}[transform canvas = {shift = {(X.north east)}}]
      \clip rectangle (\xclip, \yclip);
      \shade [right color = conecolwhite, left color = conecol, transform canvas = {rotate=-\angle}] (0,0) rectangle ++(\shadewidth,6);
    \end{scope}
    \begin{scope}[transform canvas = {shift = {(X.north west)}}]
      \clip rectangle (-\xclip, \yclip);
      \shade [right color = conecol, left color = conecolwhite, transform canvas = {rotate=\angle}] (0,0) rectangle ++(-\shadewidth,6);
    \end{scope}
    \filldraw [conecol] (X.north east) -- (cre) -- (cli) -- (X.north west) -- cycle;
    \foreach \y in {0,...,\nysteps}
    \draw [tick, shift = (origin)] ($(-\slen,\y*\ticksdist)$) -- ($(\slen,\y*\ticksdist)$);
    \draw [shift = (origin),axis] (0,0)  -- +(0,4) coordinate (ytip);
    \draw [helplines] (X.north west) -- (cli);
    \draw [helplines] (X.north east) -- ++(0,\yclip) coordinate (c1)
    (X.north east) -- (cre);
    \path (c1) ++(0,-1) coordinate (vtli)
    ++ (4,0) coordinate (vtlixshift);
    \coordinate (vtre) at (intersection cs: first line = {(vtli) -- (vtlixshift)},
    second line = {(X.north east) -- (cre)});
    \draw [widtharrow] (vtli) to node [above] {$\ v\,(t-s)$} (vtre);
    \path (vtli) ++ (-5,0) coordinate (c2);
    \coordinate (vty) at (intersection cs: first line = {(c2) -- (vtli)},
    second line = {(origin) -- (ytip)});
    \node at (vty) [anchor = east]{$t-s$};
    \node (Xlabel) [below = of innerX, Acol]{$A_X$};
    \path (Xlabel) ++(-1,0) node (Vlabel) [Vcol]{$V'$};
    \draw (cli) ++(-\shadexwidth,0) coordinate (cli2) [helplines]-- ++(0,-\yclip) coordinate (cli2b);
    \draw (cre) ++(\shadexwidth,0) [helplines] coordinate (cre2)-- ++(0,-\yclip) coordinate (cre2b);
    \coordinate (cDli) at (X.north east);
    \coordinate (cDre) at (cre2b);
    \draw [widtharrow] (cDli) to node [above] {$D$} (cDre);
  \end{tikzpicture}
  \caption{The space time cone of an observable $A_X$ in one spatial dimension. The truncation error scales exponentially in the distance $D- v\cdot(t-s)$.}
  \label{fig:lightcone}
\end{figure}
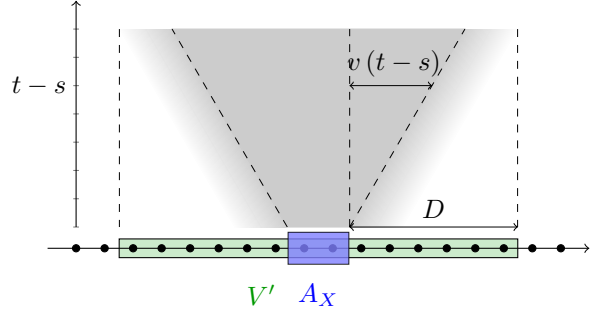

The bound from the following theorem is visualized in Fig.~\ref{fig:lightcone}. 
\begin{theorem}[Quasi-locality of local Liouvillian dynamics \cite{LRTrotter}]\label{thm:locality}
Let $\L: \RR \to \LL(\H)$ be a local Liouvillian as specified in Eqs.~\eqref{eq:localLdef} --  \eqref{eq:nndef} and let its inter\-action graph 
be of spatial dimension $\sdim$ with the constant $\sdimconst$ as defined in Eq.~\eqref{eq:sdimdef}. 
Then, for all $X\subset \Vset' \subset \Vset$ with $D \coloneqq \dist(X, \compl{\Vset'}) \geq 2\sdim -1$,
$A_X \in \B_X(\H)$, and $s\leq t$
\begin{equation}
 \norm{\Tad[\trunc{L}{V'}] s t(A_X) - \Tad[L] s t(A_X)}
 \leq \frac{2 \sdimconst}{\nn} D^{\sdim-1} \E^{v \cdot (t-s) - D} \norm{A_X},
\end{equation}
where $v= \LRv$ is the Lieb-Robinson speed. 
\end{theorem}
So, colloquially speaking, the full dynamics of local observables can be approximated with exponential accuracy by the dynamics of a sufficiently large subsystem. 
Of course, the size of the subsystem depends on the desired time span of the evolution.
In particular the locality result makes an extension of time evolution to infinitely large lattices possible, i.e., it can be employed to rigorously define the thermodynamic limit.

Theorem~\ref{thm:locality} has two further immediate physical consequences, which can be seen as an interpretation of the result.
For the rest of this section consider a lattice system with $V=\{1,\dots,N\}$ and let $\rho$ be a product state, i.e., $\rho = \bigotimes_{j=1}^{N} \rho_j$ where $\rho_j \in \mcS(\H_{\{j\}})$ for all $j$ and moreover, let $X,Y \subset V$ such that $X \cap Y = \emptyset$.
\begin{description}
\item[Suppression of correlation functions:]
  Consider two observables $A_X\in \A_X$ and $B_Y\in \A_Y$. 
  Their correlation coefficient in state $\rho \in \mcS(\H)$ is proportional to the \emph{covariance}
  \begin{equation}
    \cov_\rho(A_X,B_Y) \coloneqq \ex {A_X B_Y} \rho - \ex {A_X} \rho \ex {B_Y} \rho .
  \end{equation}
  If $\rho$ is a product state, $\cov_\rho(A_X ,B_Y) = 0$.
  Now, Theorem~\ref{thm:locality} tells us that as long as $v \cdot (t-s) \ll \dist(X,Y)/2$ the correlation coefficient of the time evolved observables will remain very small. More precisely, 
  $\cov_\rho(\Tad s t (A_X),\Tad s t (B_Y))$ is upper bounded by $\exp(v \cdot (t-s) - \dist(X,Y)/2)$ up to a constant factor. The measurement statistics of the two observables can show correlations only after the dynamics of the system had enough time to correlate the two regions $X$ and $Y$ (see Ref.~\cite{Nachtergaele2006_124} for a similar discussion in the context of Hamiltonian dynamics).
\item[Speed of information propagation:] 
Time evolution on a lattice system can also be thought of as a channel that one might want to use to send information from $X$ to $Y$ in the following way: One party encodes a message by preparing at time $s$ the part of the initial state $\rho$ in the region $X$ in a particular way, the other party tries to retrieve the message by measuring on region $Y$ at time $t$.
    Lieb-Robinson bounds can be used to show that the amount of information that can be transferred in this way in a time span $t-s$ is exponentially suppressed if $\dist(X,Y)$ is larger than $v \cdot (t-s)$. This can be made precise in the sense that the classical information capacity is exponentially small outside the cone, if the quantum many-body systems is used as a quantum channel \cite{Bravyi2006-97}. 
    In Ref.~\cite{Bloch}, the ballistic propagation of excitations and information propagation 
    constrained by Lieb-Robinson bounds  has been experimentally explored in systems of cold atoms.
\end{description}

\subsection{Classical simulation of quantum dynamics}
\label{sec:simulationofquantumdynamics}
By classical simulation of quantum dynamics we mean the calculation of expectation values of local observables $\ex{A_X} \rho (s,t)$, so that one could, for instance, plot them over time. If one tries to do that naively, i.e., by calculating the full propagator $\Tad[L] st$ on a classical computer, one quickly runs into problems even with just having enough memory to store the propagator. For instance, if one has $N$ subsystems with a local Hilbert space dimension of $2$, then to completely specify the propagator in a naive way, one needs $2^{4N}$ complex numbers.
Therefore, if one aims at classically simulating 
local observables one needs to come up with a smart simulation scheme that only deals with the information relevant for the simulation. 
We sketch two such schemes here:
\begin{description}
\item[Time evolution as (unitary) circuits:]
Here the quasi-locality Theorem~\ref{thm:locality} is of great help, since it already tells us that one can truncate the dynamics to a set $\Vset'$ containing the space time cone of the observable instead of considering the full system $V$.
The arising error is exponentially small in the distance between the space time cone and the truncation.
So the simulation cost does clearly not depend on the system size and the dynamics can hence be implemented efficiently in that.
Of course, implementing the full simulation naively on $\Vset'$ is still by far not optimal.
Famously, one can decompose the propagator $\Tad[\trunc{L}{\Vset'}]st$ into products over short time steps and strictly local propagators
, which is often called Trotter-decomposition \cite{Trotter1959}.
At the heart of this approach is the following \emph{product formula} that can be used to bound the error one makes by decomposing the propagator of a Liouvillian that is a sum of two Liouvillians $\L$ and $\K$ into the product of the propagators of these Liouvillians:
\begin{theorem}[Trotter product formula \cite{Kliesch2011-107,LRTrotter}]
  Let 
  \begin{equation}
   \L = \sum_{X \in \Eset} \L_X
  \end{equation}
  be a Liouvillian with $\L_X\in \LL_X$. Then there exist constants $b$ and $c$ that depend only on local properties of $\L$, and are in particular independent of the number of sites, 
  such that for all $X \in \Eset$ and operators $A$
  \begin{equation}
    \norm{\Tad[L]st(A) - \Tad[L]st \Tad[L-L_{\mathnormal{X}}]st(A)} \leq c (t-s)^2 \E^{b (t-s)} |\Eset| \norm{A}.
  \end{equation}
\end{theorem}
One can now decompose the time span $t-s$ into short time steps $[s_{j+1},s_j]$ and in each of these intervals approximate the propagator by a product of the strictly local propagators $\Tad[L_{\mathnormal X}]{s_{j+1}}{s_j}$ for each edge $X$ in the interaction graph of the Liouvillian. 
In other words, the full propagator can be approximated by a ``circuit" of strictly local propagators. 
The number of time steps needed to reach a simulation with total error upper bounded by $\epsilon$ is proportional to $(t-s)^2 |\Eset|^2 /\epsilon$ \cite{Kliesch2011-107}. 
Of course, the above covers Hamiltonian dynamics as a special case. However, there one would rather apply similar ideas to the time evolution operator $\exp(-\I\, (t-s) H)$ rather than the propagator. 
In a variant of this circuit description for Hamiltonian dynamics in 1D, the time evolution operator can be approximated by a circuit of constant depth and time-dependent gates \cite{QCA}.

\item[Time-dependent density-matrix renormalization group methods:] 
A similar mindset is also fundamental for the simulation of time evolution using so-called tensor network states. 
The situation is particularly clear in 1D systems with sites $V=\{1,\dots, N\}$
in pure states undergoing local Hamiltonian dynamics. If the initial state has a strong decay of initial correlations, then
the time evolution can for short times be efficiently grasped in terms of \emph{matrix-product states} (MPS) \cite{FCS,AgeMPS,MPS}. 
These are variational state vectors that are described by 
$O(d\,N D^2)$ variational parameters, where $D\in \NN$ is a refinement parameter and $d$ the dimension of the local Hilbert space. There are several variants of this approach, based on either a Trotter-decomposition \cite{Trotter1959} or a time-dependent variational principle \cite{TDVP}. 
Such schemes are subsumed under the term \emph{time-dependent density matrix renormalization group method} (t-DMRG).
At the heart of the functioning of t-DMRG is the insight that states generated by short time local Hamiltonian dynamics will have \emph{low entanglement}.
This can be formalized \cite{Eisert2006-97} in terms of so-called area laws 
\cite{Area1d,Detectability,StabilityVerstraete,Review}
that arise as a consequence of a Lieb-Robinson bound.

An area law is an upper bound on the entanglement of a state. More precisely, we say that a pure state satisfies an \emph{area law} if for any region $R \subset \Vset$
the (R\'enyi) entropy of the reduced state on $R$ can be bounded by the size of the boundary of $R$, up to a constant.
States of 1D systems satisfying an area law can be 
provably well approximated by matrix product states \cite{VerCir-MPSdim-bound}. 
Indeed, t-DMRG simulates time evolution for short times to essentially machine precision. 
For long times, the entropy will in general grow too much, as then sites are in the space time cone of too many sites of the lattice, and an 
efficient simulation in terms of matrix-product states is hence \cite{SchuchApproximation} no longer possible \cite{SchuchLongTimes,Quench}. 
That is, the power of the t-DMRG approach can be rigorously grasped in terms of Lieb-Robinson bounds. 
For 1D local Liouvillian dynamics, variants of t-DMRG have also been proposed \cite{MPO1,MPO2}, 
usually as variational principles over \emph{matrix-product operators}, the mixed state analogues of matrix-product states, 
or by means of suitable sampling employing classical stochastic processes in Hilbert space \cite{Unravelling}.
\end{description}

\subsection{Static properties derived from Lieb-Robinson bounds}

Among the most important applications of Lieb-Robinson bounds are proof techniques related to static (time independent) properties of quantum lattice systems.
Here we briefly mention some of them:
\begin{description}
\item[Clustering of correlations in Hamiltonian systems:]
One of the most relevant applications concerns the decay of correlations in the ground state of a local Hamiltonian with a spectral gap\footnote{
The spectral gap of a Hamiltonian $\Delta E$ is the difference between the ground state energy and the energy of the first exited state.
}, first shown in Refs.~\cite{Has04-LSM,Has06_clustering_gs} and further generalized in Ref.~\cite{NacSim06_clustering}.
The basic intuition underlying this intricate insight is that the spectral gap $\Delta E$ essentially defines a time scale in the system, which in turn can be related to a length scale.
\begin{theorem}[Clustering of correlations in unique ground states \cite{Has06_clustering_gs,NacSim09_review}]
  Let $H \in \A(\H)$ be a local Hamiltonian with a unique ground state $\psi$ and a spectral gap $\Delta E>0$
  and $X,Y \subset \Vset$. 
  Then, for every 
  $A_X \in \B_X(\H)$ and $B_Y \in \B_Y(\H)$
\begin{equation}\label{eq:Clustering}
  \left| \cov_\psi (A_X, B_Y) \right|
  \leq C \norm{A_X}\norm{B_Y} \E^{-\mu \dist(X,Y)}.
\end{equation}
$C$ and $\mu$ are constants both depending on $\Delta E$. Moreover, $C$ depends on the lattice geometry and the smaller of the surface areas of $X$ and $Y$, and $\mu$ depends on the Lieb-Robinson speed.
\end{theorem}
The proof of this statement confirmed a long-standing conjecture in condensed-matter physics, that gapped Hamiltonian systems have exponentially clustering correlations in the ground state.

\item[Clustering of correlations in Liouvillian systems:]
A similar intuition actually holds true for Liouvillian systems, where the role of the ground state of Hamiltonian systems is taken over by the stationary state.
Clustering of correlations in local Liouvillian systems has first been considered in Ref.~\cite{Poulin2010-104} and has been made rigorous and largely generalized in Ref.~\cite{Kastoryano2013}:
If a local Liouvillian is primitive (that is, if its stationary state has full rank) and has a spectral gap which is independent of the system size, then correlation functions between local observables again decay exponentially as a function of the distance between their supports. 

\item[Area laws of ground states of gapped Hamiltonians:] 
It has been shown using Lieb-Robinson bounds that ground states of 1D local Hamiltonian systems with 
spectral gap $\Delta E>0$ always satisfy an area law for the R\'enyi entropies 
(for a review, see Ref.~\cite{Review}).
This result has since been tightened \cite{Detectability} and area laws have also been shown for some instances of gapped higher-dimensional Hamiltonian systems \cite{HD}. It has also been shown that in 1D exponential clustering of correlations already implies an area law \cite{BraHor13}.
For local Liouvillians, general area laws (in terms of entropic measures suitable for mixed states) can be derived for stationary states \cite{Kastoryano2013}, again using Lieb-Robinson bounds.

\item[Approximating 1D ground states of gapped Hamiltonians with MPS:]  
Since ground states of any 1D local Hamiltonian with a spectral gap $\Delta E>0$ satisfy an area law for R\'enyi entropies they can be approximated \cite{VerCir-MPSdim-bound} by matrix product states (MPS) in polynomial time \cite{LanVazVid13}. 
This is used by the static \emph{density-matrix renormalization group method} (DMRG) \cite{Schollwoeck2005} (see also the chapter of Ors Legeza, Thorsten Rohwedder and Reinhold Schneider) for simulating ground state properties \cite{AgeMPS}, which has led to a wealth of novel insights in condensed matter physics.

\item[Higher-dimensional Lieb-Schultz-Mattis theorems:] 
The Lieb-Schultz-Mattis theorem \cite{LSM1,LSM2} is an upper bound on the spectral gap of certain one-dimensional quantum spin systems.
Using Lieb-Robinson bounds, a higher-dimensional Lieb-Schultz-Matthis theorem has been proven in Refs.~\cite{Has04-LSM,LSM3}.

\item[Stability and further properties of ground states:]
Lieb-Robinson bounds are one of the pillars of the formalism grasping the stability of ground states of 
a certain class of Hamiltonians (frustration-free Hamiltonians satisfying certain topological order conditions) under local perturbations.
This has developed into a field of research in its own right, and we merely touch upon the topic here. Starting point is the concept of \emph{quasi-adiabatic continuation} \cite{Wen}, which is a tool to connect dynamical properties of a Hamiltonian to static ones and relies on Lieb-Robinson bounds. 
Importantly, quasi-adiabatic continuation is a cornerstone of the proof of the stability of topological order under local perturbations \cite{TopologicalStability} and related proofs of the stability of the spectral gap, of frustration-free Hamiltonians under general, quasi-local perturbations \cite{Justyna}. 
With similar tools, the stability of the area law for the entanglement entropy of the ground state can be proven \cite{StabilityAreaLaw,StabilityVerstraete}.

\item[Stability of stationary states:]
Inspired by the stability results on Hamiltonian ground states, Lieb-Robinson bounds have also been used to prove the stability of stationary states of certain local Liouvillians \cite{PerezGarcia2013,Kastoryano2013}.

\item[Structure of elementary excited states:] 
The structure of elementary excited states has been explored using Lieb-Robinson bounds, which can be 
approximated by superimposing ground states to which local operators have been applied \cite{Excited}.
\end{description}

\section{Fermionic Hamiltonians}
\label{sec:jordanwignertransformation}
While Lieb-Robinson bounds are usually stated for spin lattice system, they also hold for systems of fermions on a lattice. 
The situation is particularly simple for 1D systems with nearest neighbor coupling only, since in that case the Jordan-Wigner transform can be applied. In this section we first state a fermionic Lieb-Robinson bound and then introduce the Jordan-Wigner transform. 

Again, as with spin lattice systems, we have an interaction (hyper)graph $(\Vset, \Eset)$ but now work in the picture of second quantization, i.e., operators are given in terms of the fermionic creation and annhilation operators $f_j$ and $f\ad_k$ for $j,k \in \Vset$. These fermionic operators satisfy
\begin{equation}
	\{ f_j ,f_k^\dagger \} = \delta_{j,k}, 
\end{equation}
where $\{A,B\}\coloneqq AB + BA$ is the anti-commutator. 
According to the \emph{fermion number parity superselection rule} only observables that are even polynomials in the fermionic operators can occur in nature. A polynomial of fermionic operators is called \emph{even} if it can be written as a linear combination of monomials, where each monomial is a product of an even number of fermionic operators from $f_j$ and $f\ad_k$. 
Correspondingly, we denote the algebra of the parity preserving observables acting on a region $X \subset \Vset$ by $\G_X$ for short. Now one can prove 
a fermionic Lieb-Robinson bound in the same way as Theorem~\ref{thm:LR} is proven:
\begin{theorem}[Fermionic Lieb-Robinson bound]
Let 
\begin{align}
 H = \sum_{X \in \Eset} H_X
\end{align}
be a local time-dependent Hamiltonian with $H_X : \RR \to \G_X$ and $\norm{H_X(r)} \leq b$ for all $X \in \Eset$ and $r \in \RR$, $\tau$ its propagator, and $\nn$ the maximum number of nearest neighbors as defined in Eq.~\eqref{eq:nndef}. 
Then, for every $A_X \in \G_X$, $B_Y \in \G_Y$ and $s , t \in \RR$,
\begin{equation}\label{eq:fermionic_LRbound}
\norm{[B_Y, \Tad s t (A_X)]} \leq C \norm{B_Y} \norm{A_X} \E^{v |t-s| - \dist(X,Y)},
\end{equation}
where $v = \LRv$ and $C$ is some constant depending polynomially on the size of the smaller of the two sets $X$ and $Y$. 
\end{theorem}
For the unphysical case where $B_Y$ and $A_X$ are observables that are odd polynomials in the fermionic operators one can still prove a similar Lieb-Robinson bound for the anti-commutator, providing a relevant proof-tool \cite{Has04-fermionicLRbound}.

For the case of 1D systems with nearest neighbor interactions only, the analogy between fermionic and spin systems is even stronger in the sense that such systems can be mapped to each other by the Jordan-Wigner transform \cite{JordanWigner}. 
Note that a higher-dimensional variant has also been developed \cite{Verstraete05}. 

Consider a one-dimensional lattice with vertices $V = \{1,\hdots,N\}$.
The Hilbert space of the spin-$1/2$ model on $V$ is given by 
$\H \coloneqq \bigotimes_{j \in V}\H_{j}$ with $\H_{j} \cong \C^2$.
We denote by $X_j,Y_j,Z_j\in \A_{\{j\}}$ the Pauli operators acting on site $j$ of the spin chain.
Then the Jordan-Wigner-Transformation is given by
\begin{align}
  f_j +f_j\ad &= w_{2j-1} \coloneqq X_j \prod_{j'<j} Z_{j'}\\
  \I f_j - \I f_j\ad &= w_{2j} \coloneqq Y_j \prod_{j'<j} Z_{j'} ,
\end{align}
where the $(w_j)_{j=1}^{2N}$ are called \emph{Majorana operators}.
The Majorana operators satisfy the anti-commutation relation $\{w_j,w_k\} = 2 \delta_{j,k}$.  
It can be verified with elementary calculations that
\begin{align}
  f_j &= \kw 2 \left( w_{2j-1} - \I w_{2j} \right), \\
  f_j\ad f_j &= \kw 2 \left( 1- \I w_{2j-1}w_{2j} \right) ,
 \end{align}
as well as 
\begin{align}
  Z_j &= -\I w_{2j-1} w_{2j} = 2 f_j\ad f_j -1 ,\,\\
  X_j &= w_{2j-1} \prod_{j'<j} Z_{j'},\, \qquad
  Y_j = w_{2j} \prod_{j'<j} Z_{j'} ,
\end{align}
and 
\begin{equation}\label{eq:jwlocality}
  \forall j\leq k:\quad f_j\ad f_{k} = \frac{1}{4} S^+_j (\prod_{j\leq j'<k} Z_{j'}) S^-_k ,
\quad \text{where }\quad  
S^\pm_j \coloneqq X_j\pm\I Y_{j} .
\end{equation}
Most importantly, as can be seen from Eq.~\eqref{eq:jwlocality}, the Jordan-Wigner-Trans\-formation preserves locality in the sense that a one-dimensional fermionic Hamiltonian with nearest neighbor or short range hopping and short range density-density interactions is mapped to a spin chain Hamiltonian with only short range interactions.


\section{Conclusion}
We have reviewed the Heisenberg picture for time-dependent Liouvillian dynamics in spin lattice systems. For this setting we have stated a Lieb-Robinson bound. Such bounds give rise to a plethora of statements about locally interacting systems which we have reviewed subsequently. Finally, we have explained the relevance for fermionic systems. 
We hope that this text serves as an introduction to Liouvillian dynamics on spin lattice systems and 
provides an overview of important consequences of Lieb-Robinson bounds. 

\section*{Acknowledgments}
We thank Earl T.\ Campbell, Mathis Friesdorf and Albert H.\ Werner for comments. 
We acknowledge support from the EU (Q-Essence, Raquel), the BMBF (QuOReP), the ERC (Taq), and the Studienstiftung des Deutschen Volkes.


\end{document}